\documentclass[10pt,onecolumn]{article} 
\usepackage[utf8]{inputenc}
\usepackage{style_arxiv}
\usepackage{amsmath}
\usepackage{amsfonts}
\usepackage{amssymb}
\usepackage{url}
\usepackage[symbol*]{footmisc}
\setfnsymbol{wiley}
\renewcommand{\vec}[1]{\mathbf{#1}}
\ifx\pdfoutput\undefined
	\usepackage[dvips]{graphicx}
\else
	\usepackage[pdftex]{graphicx}
	\usepackage{epstopdf,subfigure}
\fi
\graphicspath{./}
\newcounter{subfig}
\begin{document}

\title{Simulation electronic thermes of two atoms molecules}%Modeling

\author{Vladimir.~P.~Koshcheev$^{1,}$\email{koshcheev1@yandex.ru}~, Yuriy.~N.~Shtanov$^{2,}$\email{yuran1987@mail.ru}}

\affiliation{\small{\textit{$^1$Moscow Aviation Institute (National Research University),}}\\ 
\small{\textit{Strela Branch, Moscow oblast, Zhukovskii, 140180 Russia}}\\
\small{\textit{$^2$Tyumen Industrial University, Surgut Branch, Surgut, 628404 Russia}}}

\preparedfor{}

\maketitle
\thispagestyle{empty}

\begin{abstract}
In the first order of the perturbation theory, the correction to the electronic terms of a diatomic molecule is calculated taking into account the Pauli principle.
\end{abstract}
\par 
\\
\\
\\
\section{Introduction}

A new approach to calculating the potential energy of interaction of two atoms \cite{Koshcheev2018,Koshcheev2020} satisfactorily describes the experimental results for atoms of noble gases, if the atomic form factor of an isolated atom is chosen in the Molière approximation. In this publication, the electronic terms of the diatomic molecule (dimer) $ {\rm HF} $ and $ {\rm HNe} $ will be constructed in the first order of the perturbation theory.

\section{Model}
\par The diatomic molecule will be described using the stationary Schrödinger equation
\begin{equation} \label{eq1} 
H{\rm \psi }=E{\rm \psi }.  
\end{equation} 

\par We represent the Hamiltonian of the equation \eqref{eq1} in the form
\begin{equation} \label{eq2} 
H=H^{0}+U; 
\end{equation} 
\[U=\frac{Z_{1} Z_{2} e^{2} }{\left|\vec{r}_1-\vec{r}_2\right|} +\sum _{j_{1} =1}^{Z_{1} }\sum _{j_{2} =1}^{Z_{2} }\frac{e^2}{\left|\vec{r}_{1}+\vec{r}_{1j_{1}}-\vec{r}_{2}-\vec{r}_{2j_{2}} \right|}-\sum _{j_{1}=1}^{Z_{1}}\frac{Z_{2} e^{2} }{\left|\vec{r}_{1}+\vec{r}_{1j_{1}}-\vec{r}_{2} \right|}  -\sum _{j_{2}=1}^{Z_{2}}\frac{Z_{1} e^{2} }{\left|\vec{r}_{1}-\vec{r}_{2}-\vec{r}_{2j_{2}} \right|},\] 
where $U$--potential energy of interaction of two atoms; $\vec{r}_1$ and $\vec{r}_2$ -- coordinates of the first and second atomic nucleus; $\vec{r}_1+\vec{r}_{1j_{1}}$ and $\vec{r}_{2}+\vec{r}_{2j_{2}}$ -- coordinates of $j_{1}$-th and $ j_{2}$-th electrons the first and second atom, respectively.

\par The solution of the equation \eqref{eq1} with the Hamiltonian \eqref{eq2} will be sought using the perturbation theory
\[{\rm \psi }={\rm \psi }^{0} +{\rm \psi }^{1} +\ldots \] 
\[E=E^{0} +E^{1} +\ldots \] 

\par The electronic terms of a diatomic molecule will be sought in the first order of the perturbation theory
\begin{equation} \label{eq3} 
E^{1} \left(r\right)=\left\langle \left. {\rm \psi }^{0} \right|\left. U\right|{\rm \psi }^{0} \right\rangle ,  
\end{equation} 
where $r=\left|\vec{r}_{1}-\vec{r}_{2}\right|$; angle brackets $\left\langle \ldots \right\rangle $ were introduced by Dirac \cite{dirac2013principles}. 

\par We represent the Hamiltonian $H^0$ in the form
\[H^0=H_1^0+H_2^0,\] 
where $H_{i}^{0}$--the Hamiltonian of the \textit{i}th atom; \textit{i}=1,2.

\par The solution to the Schrödinger equation
\[H^{0}{\rm \psi}^{0}=E^{0}{\rm \psi}^{0},\] 
will be sought in the form
\[{\rm \psi}^{0}={\rm \psi}_{1}^{0}{\rm \psi}_{2}^{0},\] 
\[E^{0}=E_{1}^{0}+E_{2}^{0},\] 
where the Schrödinger equation for the \textit{i}th isolated atom has the form
\begin{equation} \label{eq4} 
H_{i}^{0}{\rm \psi}_{i}^{0}=E_{i}^{0}{\rm \psi}_{i}^{0},  
\end{equation} 
where ${\rm \psi}_{i}^{0}={\rm \psi}_{i}^{0}\left(\vec{r}_{i1}, \vec{r}_{i2},\ldots, \vec{r}_{iZ_{i}} \right)$.

\par It is known \cite{Fock2007} that using the variational principle from the stationary Schrödinger equation \eqref{eq4} one can construct the Hartree-Fock equation. Hydrogen-like wave functions that approximate the solution of the Hartree-Fock equation for an isolated atom are presented in \cite{CLEMENTI1974177}
\begin{equation} \label{eq5} 
{\rm \psi }_{i}^{0} =\vartheta _{i1} \left(\vec{r}_{i1} \right)\vartheta _{i2} \left(\vec{r}_{i2} \right)\ldots \vartheta _{iZ_{i} } \left(\vec{r}_{iZ_{i} } \right),  
\end{equation} 
where $\vartheta_{ij}=\vartheta _{ij} \left(\vec{r}_{ij} \right)$--hydrogen-like wave functions that form an orthonormal system. 

\par Using the formulas \eqref{eq2} and \eqref{eq5}, we calculate \eqref{eq3} following \cite{Koshcheev2018}
\begin{equation} \label{eq6} 
E^{1} \left(r\right)=\int E^{1} \left(k\right)\exp \left(i\vec{k}\vec{r}\right) \frac{{\rm d}^{3} \vec{k}}{\left(2{\rm \pi }\right)^{3} } ;  
\end{equation} 
\[E^{1} \left(k\right)=\frac{4{\rm \pi }Z_{1} Z_{2} e^{2} }{k^{2} } \left[1-\frac{F_{1} \left(k\right)}{Z_{1} } \right]\left[1-\frac{F_{2} \left(k\right)}{Z_{2} } \right].\] 

The formula \eqref{eq6} does not take into account the Pauli principle between the electrons of the first and second atoms. In \cite{Koshcheev2018}, by analogy with (see, for example, \cite{pitaevskii2012physical}), it was proposed to take into account the Pauli principle using the factor
\begin{equation} \label{eq7} 
P\left(k\right)=\left[1-\frac{F_{1} \left(k\right)}{Z_{1} } \right]\left[1-\frac{F_{2} \left(k\right)}{Z_{2} } \right].  
\end{equation} 

\par The value $F_{i}(k)/Z_{i}$ is the Fourier component of the atomic electron distribution plane, which is normalized to unity. As a result, we obtain an expression for the electronic term (potential energy of interaction of two atoms $U(r)=E_{p}^{1}(r)+\ldots$) taking into account the Pauli principle in the form
\begin{equation} \label{eq8} 
E_{p}^{1} \left(r\right)=\int E^{1} \left(k\right)P\left(k\right) \exp \left(i\vec{k}\vec{r}\right)\frac{{\rm d}^{3} \vec{k}}{\left(2{\rm \pi }\right)^{3} } .  
\end{equation} 

\par The condition for the applicability of the correction in the first order of perturbation theory to the energy of the system in the unperturbed state has the form
\[\left|E_{p}^{1} \left(r\right)\right|\ll \left|E^{0} \right|,\] 
where $E^{0} =E_{1}^{0} +E_{2}^{0} $;the energies $ E_{i}^{0}$ are also presented in \cite{CLEMENTI1974177} together with hydrogen-like wave functions that approximate the solution of the Hartree-Fock equation for an isolated atom.

\section{Calculation results and their discussion}

\par The potential energy of interaction of two atoms was simulated when the form factor of the first atom was chosen in the Molière approximation
\begin{equation} \label{eq9} 
U\left(r\right)\approx \int \frac{4{\rm \pi }Z_{1} Z_{2} e^{2} }{k^{2} } \left[1-\frac{F_{1} \left(k\right)}{Z_{1} } \right]^{2} \left[1-\frac{F_{2} \left(k\right)}{Z_{2} } \right]^{2}  \exp \left(i\vec{k}\vec{r}\right)\frac{{\rm d}^{3} \vec{k}}{\left(2{\rm \pi }\right)^{3} } ; 
\end{equation} 
\begin{equation} \label{eq10} 
F_{1} \left(k\right)=Z_{1} \sum _{i=1}^{3}\frac{{\rm \alpha }_{i} \left({{\rm \beta }_{i} \mathord{\left/ {\vphantom {{\rm \beta }_{i}  a_{1} }} \right. \kern-\nulldelimiterspace} a_{1} } \right)^{2} }{k^{2} +\left({{\rm \beta }_{i} \mathord{\left/ {\vphantom {{\rm \beta }_{i}  a_{1} }} \right. \kern-\nulldelimiterspace} a_{1} } \right)^{2} }  ,  
\end{equation} 
where $\alpha_i,\beta_i$-- Molière approximation coefficients \cite{Moliere1947TheorieDS}; $a_1\approx 0.88534 a_0 Z_1^{-1/3}$; $a_0=0.529\AngM$.

\par Atomic form factor $F_2(k)$ was calculated using the wave function of the hydrogen atom in the $1s$-state 
\begin{equation} \label{eq11} 
F_{2} \left(k\right)=\frac{16}{\left(4+a_{2}^{2} k^{2} \right)^{2} } ,  
\end{equation} 
where $a_2=a_0/Z_2\approx 0.529\AngM/Z_2$ -- shielding length of a hydrogen atom. 

\par We calculate the expressions \eqref{eq9} using the formulas \eqref{eq10} and \eqref{eq11}
\begin{equation} \label{eq12} 
\begin{array}{l} {U\left(r\right)\approx \frac{18e^{2} }{r} \left[\exp\left(-\frac{14.0968r}{a_{0} } \right)\left(0.0973-0.0352\frac{r}{a_{0} } \right)+\right. } \\ {+\exp\left(-\frac{2.8194r}{a_{0} } \right)\left(-0.03236-0.0001451\frac{r}{a_{0} } \right)+} \\ {+\exp\left(-\frac{0.7048r}{a_{0} } \right)\left(0.0182-0.00199\frac{r}{a_{0} } \right)+} \\ {+\exp\left(-\frac{2r}{a_{0} } \right)\left. \left(0.416838-0.62409\frac{r}{a_{0} } +0.0846\frac{r^{2} }{a_{0}^{2} } -0.002121\frac{r^{3} }{a_{0}^{3} } \right)\right].} \end{array} 
\end{equation} 

\par Numerical calculations were performed using the program \cite{bib:KoscheevSvidetelstvo}. Figure 1 shows the graphs of the potential energy of interaction of atoms in the ${\rm HF}$ and $ {\rm HNe} $ molecules. Since the atomic form factor for fluorine and neon atoms was chosen in the Molière approximation, the results of calculating the potential interaction energy of atoms in ${\rm HF}$ and ${\rm HNe}$ molecules differ slightly from each other. 
It can be seen that the far potential minimum in Fig.1c, which corresponds to the area of action of van der Waals forces, is $10^4\div 10^5$ times less deep than the near potential minimum in Fig.1a. 
\par Figure 2 shows satisfactory agreement of the graph of the potential energy of interaction of atoms in the ${\rm HF}$ molecule in comparison with the experimental data presented in \cite{Fan2014}. In \cite{Koshcheev2020}, satisfactory agreement with the experiment was demonstrated for the plot of the potential energy of interaction of two neon atoms in the region of action of van der Waals forces. The form factors of neon atoms in \cite{Koshcheev2020} were taken in the Molière approximation. The agreement between the calculation results and experiment can be improved if the form factors of fluorine and neon atoms are constructed using hydrogen-like wave functions \cite{CLEMENTI1974177}, which approximate the solution of the Hartree-Fock equation for an isolated atom. It can be seen that for the formation of the $ {\rm HNe} $ molecule when hydrogen and neon atoms approach each other, it is necessary to overcome a potential barrier with a height of about $ 0.04$ electronvolt, as can be seen in Fig. 1b. At low temperatures, the formation of ${\rm HF} $ and ${\rm HNe}$ dimers in the region of action of van der Waals forces is possible. The depth of the potential well is $ 10^{-4}\div 10^{-5} $ electronvolts, and the distance between the atoms is $5\div 6 \AngM$ as seen in Figure 1c.

\section{The Acknowledgements}
\par The reported study was funded by RFBR, project number 20-07-00236 a.

\bibliographystyle{unsrturl}%unsrt
\bibliography{refs}

\newpage
\section*{\center List of figures}
\par \textbf{Fig. 1}. Potential energy of interaction of atoms in molecules $\rm HF$(solid line) and $\rm HNe$ (dashed line) at different scales: a)$r/a_0\in[0;3]$, b)$r/a_0\in[3;8]$, c)$r/a_0\in[8;15]$.
\par \textbf{Fig. 2}. The potential interaction energy $U(r)$, which is calculated by formula (12) (solid line), in comparison with the experimental result, which is presented in \cite{Fan2014} (circles), and the Morse potential \cite{Fan2014} (dashed line) for the $\rm HF$ molecule.

\newpage
\begin{figure}[!b]
  \begin{center}
    \includegraphics[width=\textwidth]{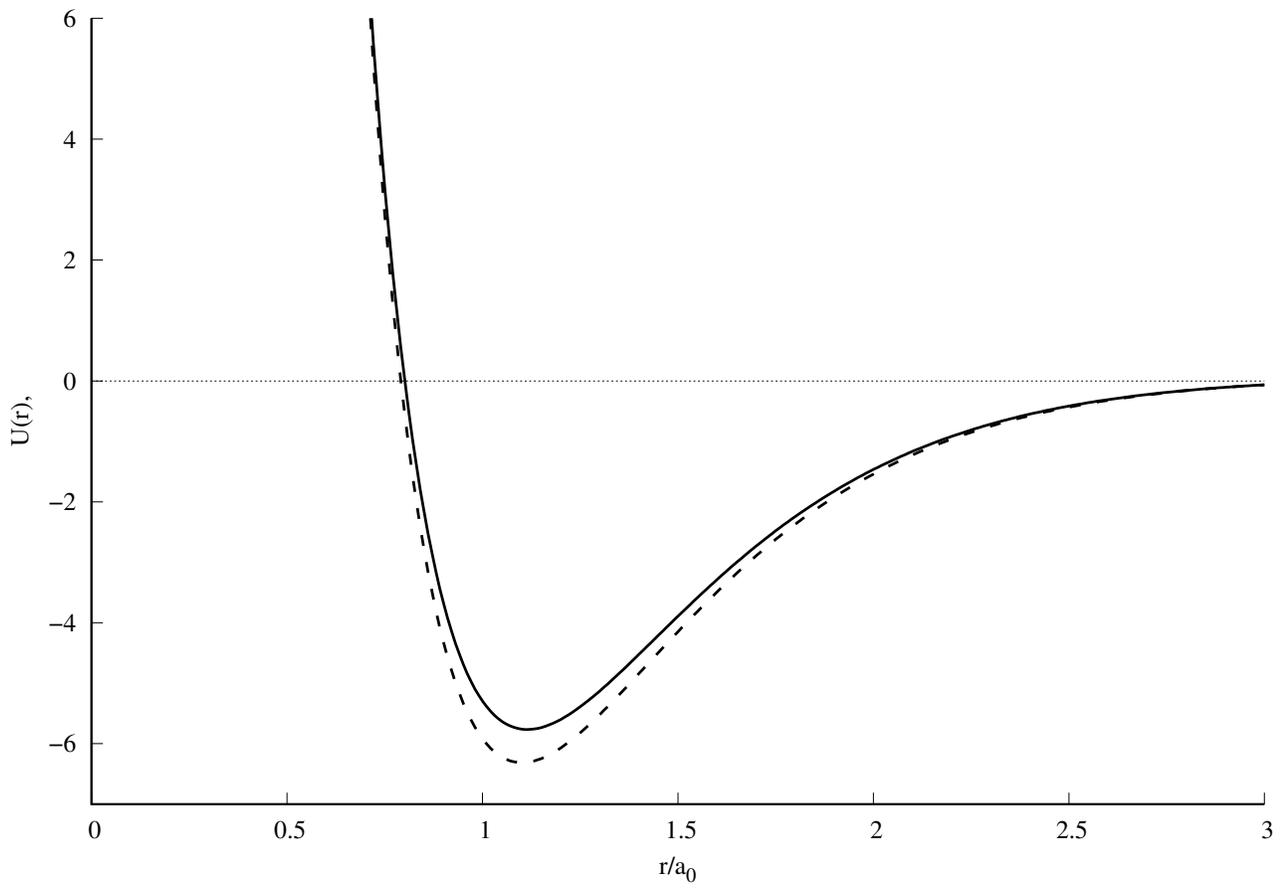}
  \end{center}

  \caption{\small Potential energy of interaction of atoms in molecules $\rm HF$(solid line) and $\rm HNe$ (dashed line) at $r/a_0\in[0;3]$}
  \label{fig1a}
  \addtocounter{subfig}{1}
  \setcounter{figure}{0}
\end{figure}

\newpage
\begin{figure}[!b]
  \begin{center}
  	\includegraphics[width=\textwidth]{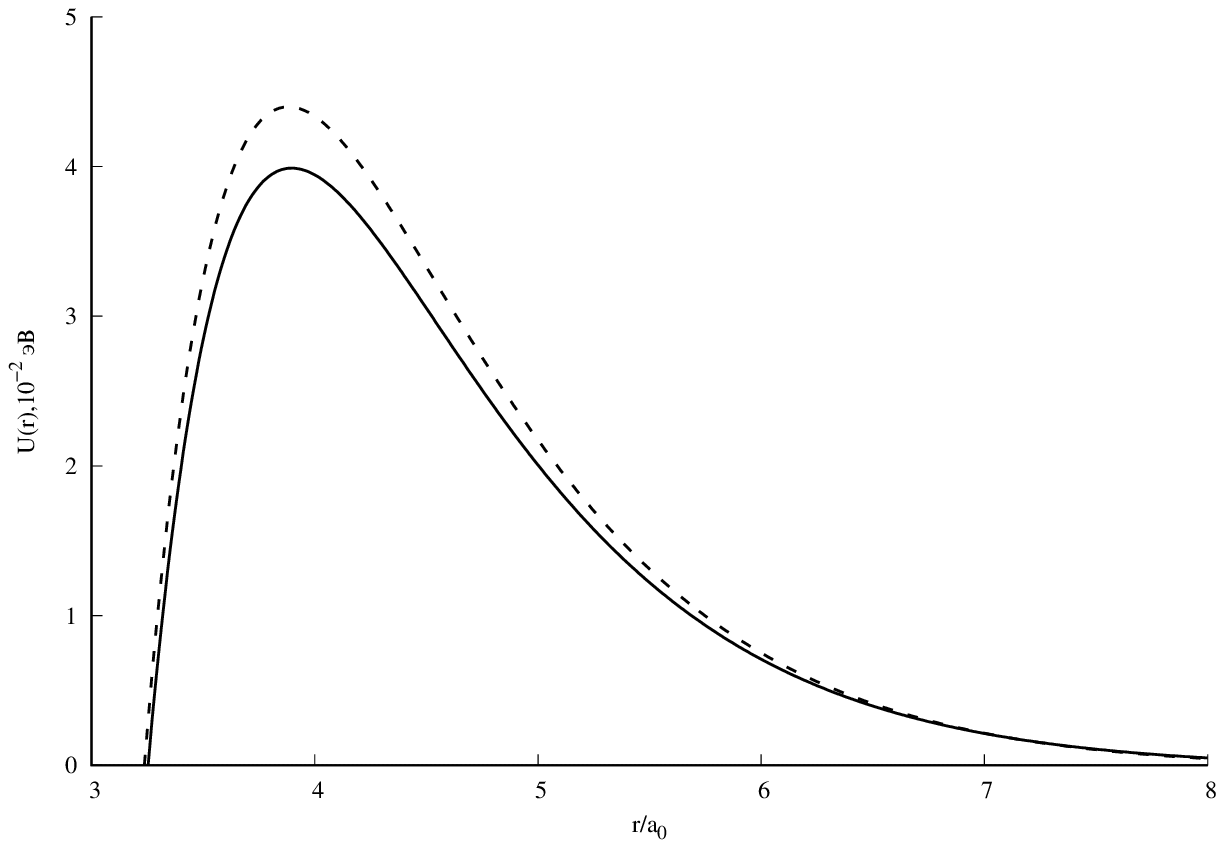}
  \end{center}

  \caption{\small Potential energy of interaction of atoms in molecules $\rm HF$(solid line) and $\rm HNe$ (dashed line) at $r/a_0\in[3;8]$}
  \label{fig1b}
  \addtocounter{subfig}{1}
  \setcounter{figure}{0}
\end{figure}

\newpage
\begin{figure}[!b]
  \begin{center}
  	\includegraphics[width=\textwidth]{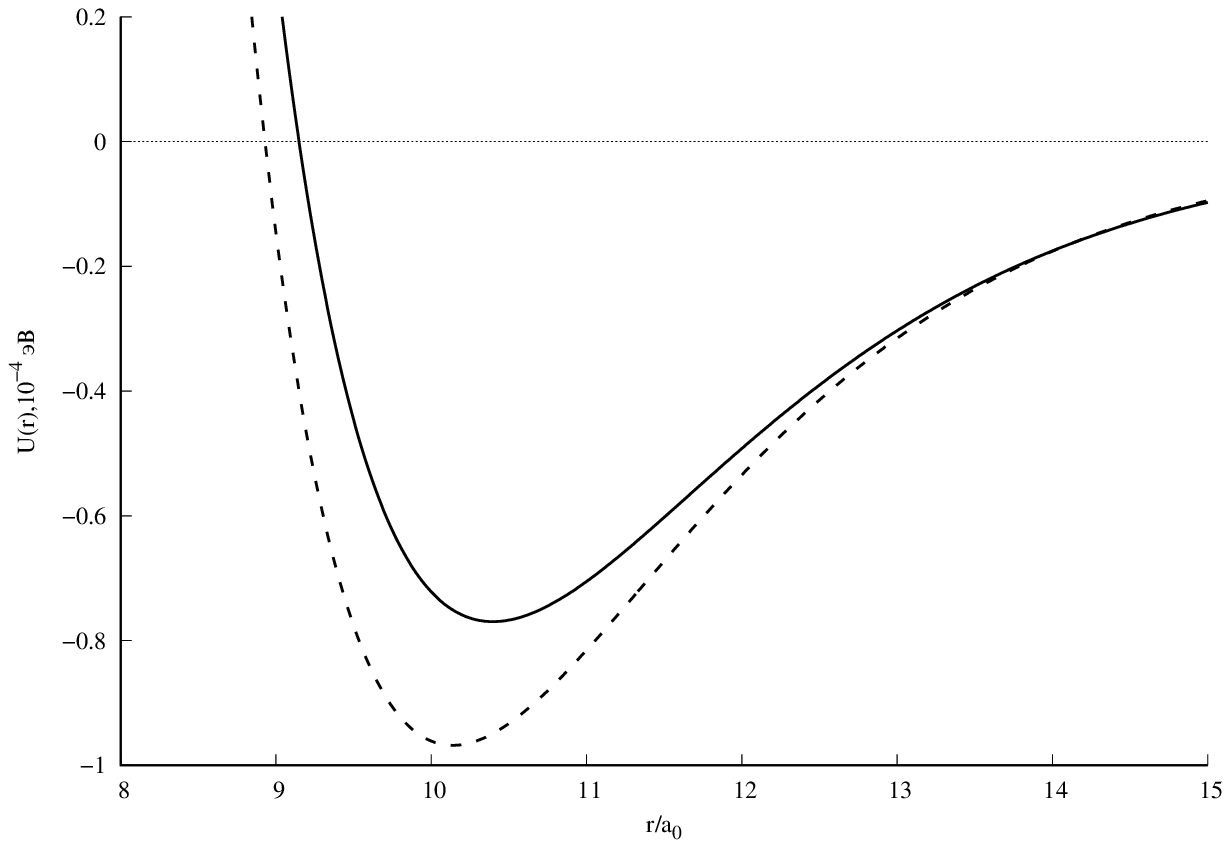}
  \end{center}

  \caption{\small Potential energy of interaction of atoms in molecules $\rm HF$(solid line) and $\rm HNe$ (dashed line) at $r/a_0\in[8;15]$}
  \label{fig1c}
  \setcounter{subfig}{0}
\end{figure}

\newpage
\begin{figure}[!b]
  \begin{center}
  	\includegraphics[width=\textwidth]{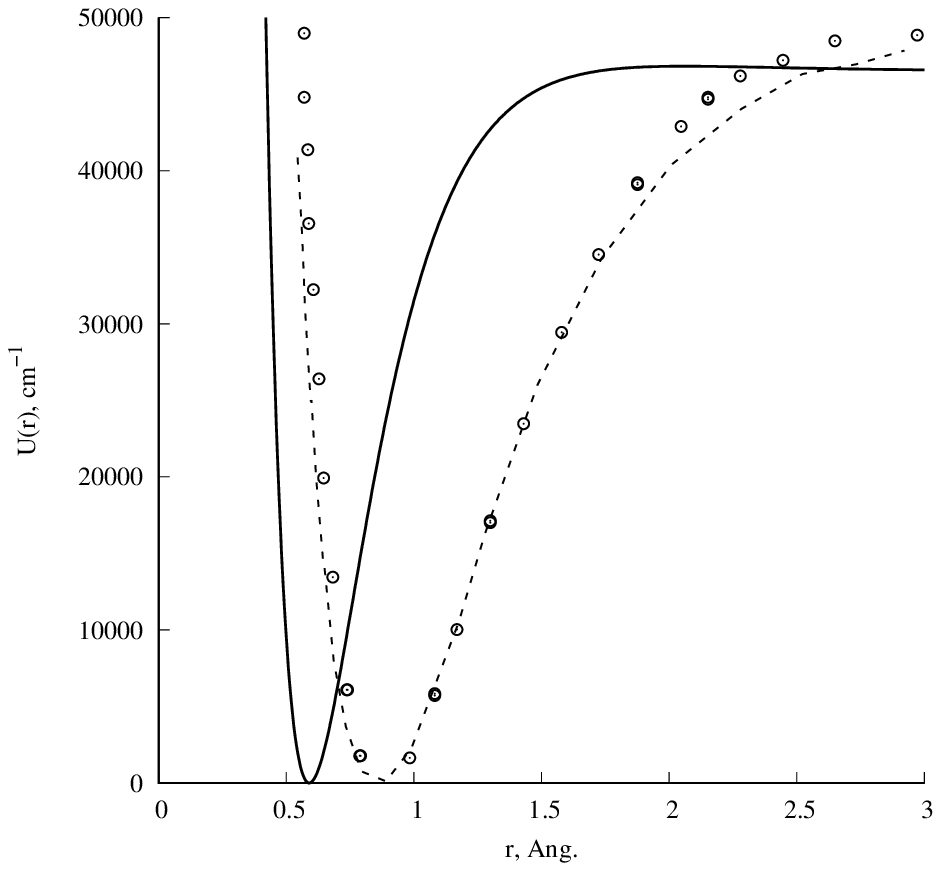}
  \end{center}

  \caption{\small The potential interaction energy $U(r)$, which is calculated by formula (12) (solid line), in comparison with the experimental result, which is presented in \cite{Fan2014} (circles), and the Morse potential \cite{Fan2014} (dashed line) for the $\rm HF$ molecule.}
  \label{fig2}
\end{figure}

\end{document}